\newcount\mgnf\newcount\tipi\newcount\tipoformule
\newcount\aux\newcount\casa\newcount\piepagina
 
\mgnf=0
\aux=0
\tipoformule=0
\piepagina=1
 
\def\9#1{\ifnum\aux=1#1\else\relax\fi}
\ifnum\piepagina=0 \footline={\rlap{\hbox{\copy200}\
$\st[\number\pageno]$}\hss\tenrm \foglio\hss}\fi \ifnum\piepagina=1
\footline={\rlap{\hbox{\copy200}} \hss\tenrm \folio\hss}\fi
\ifnum\piepagina=2\footline{\hss\tenrm\folio\hss}\fi
\newcount\referenze \referenze=0 \ifnum\referenze=0
\openout7=\jobname.b\def\[#1]{[#1]\write7{[#1]}}\fi \ifnum\referenze=1
\input \jobname.bb \fi \casa=0
 
\ifnum\mgnf=0 \magnification=\magstep0
\hsize=12truecm\vsize=19.5truecm \parindent=4.pt\fi
\ifnum\mgnf=1 \magnification=\magstep1
\hsize=16.0truecm\vsize=22.5truecm\baselineskip14pt\vglue6.3truecm
\overfullrule=0pt \parindent=4.pt\fi
 
\let\a=\alpha  \let\g=\gamma 
\let\e=\varepsilon  \let\h=\eta
  \let\m=\mu \let\n=\nu
 \let\p=\pi \let\r=\rho \let\s=\sigma 
   \let\o=\omega
  \let\D=\Delta 
 \let\P=\Pi  
 \let\O=\Omega 
{\count255=\time\divide\count255 by 60 \xdef\oramin{\number\count255}
\multiply\count255 by-60\advance\count255 by\time
\xdef\oramin{\oramin:\ifnum\count255<10 0\fi\the\count255}}
\def\ora{\oramin } \def\data{\number\day/\ifcase\month\or gennaio \or
febbraio \or marzo \or aprile \or maggio \or giugno \or luglio \or
agosto \or settembre \or ottobre \or novembre \or dicembre
\fi/\number\year;\ \ora} \setbox200\hbox{$\scriptscriptstyle \data $}
\newcount\pgn \pgn=1 \def\foglio{\number\numsec:\number\pgn
\global\advance\pgn by 1} \def\foglioa{A\number\numsec:\number\pgn
\global\advance\pgn by 1}
\global\newcount\numsec\global\newcount\numfor \global\newcount\numfig
\gdef\profonditastruttura{\dp\strutbox}
\def\senondefinito#1{\expandafter\ifx\csname#1\endcsname\relax}
\def\SIA #1,#2,#3 {\senondefinito{#1#2} \expandafter\xdef\csname
#1#2\endcsname{#3} \else \write16{???? ma #1,#2 e' gia' stato definito
!!!!} \fi} \def\etichetta(#1){(\veroparagrafo.\veraformula) \SIA
e,#1,(\veroparagrafo.\veraformula) \global\advance\numfor by 1
\9{\write15{\string\FU (#1){\equ(#1)}}} \9{ \write16{ EQ \equ(#1) == #1
}}} \def \FU(#1)#2{\SIA fu,#1,#2 }
\def\etichettaa(#1){(A\veroparagrafo.\veraformula) \SIA
e,#1,(A\veroparagrafo.\veraformula) \global\advance\numfor by 1
\9{\write15{\string\FU (#1){\equ(#1)}}} \9{ \write16{ EQ \equ(#1) == #1
}}} \def\getichetta(#1){Fig.  \verafigura \SIA e,#1,{\verafigura}
\global\advance\numfig by 1 \9{\write15{\string\FU (#1){\equ(#1)}}} \9{
\write16{ Fig.  \equ(#1) ha simbolo #1 }}} \newdimen\gwidth \def\BOZZA{
\def\alato(##1){ {\vtop to \profonditastruttura{\baselineskip
\profonditastruttura\vss
\rlap{\kern-\hsize\kern-1.2truecm{$\scriptstyle##1$}}}}}
\def\galato(##1){ \gwidth=\hsize \divide\gwidth by 2 {\vtop to
\profonditastruttura{\baselineskip \profonditastruttura\vss
\rlap{\kern-\gwidth\kern-1.2truecm{$\scriptstyle##1$}}}}} }
\def\alato(#1){} \def\galato(#1){}
\def\veroparagrafo{\number\numsec}\def\veraformula{\number\numfor}
\def\verafigura{\number\numfig}
\def\geq(#1){\getichetta(#1)\galato(#1)}
\def\Eq(#1){\eqno{\etichetta(#1)\alato(#1)}}
\def\eq(#1){\etichetta(#1)\alato(#1)}
\def\Eqa(#1){\eqno{\etichettaa(#1)\alato(#1)}}
\def\eqa(#1){\etichettaa(#1)\alato(#1)}
\def\eqv(#1){\senondefinito{fu#1}$\clubsuit$#1\write16{No translation
for #1} \else\csname fu#1\endcsname\fi}
\def\equ(#1){\senondefinito{e#1}\eqv(#1)\else\csname e#1\endcsname\fi}
\openin13=#1.aux \ifeof13 \relax \else \input #1.aux \closein13\fi
\openin14=\jobname.aux \ifeof14 \relax \else \input \jobname.aux
\closein14 \fi \9{\openout15=\jobname.aux} \newskip\ttglue
\ifnum\casa=1
\font\dodiciit=cmr12
\font\titolo=cmr12 \else
\font\dodiciit=cmti12
\font\titolo=cmbx12 scaled \magstep2 \fi
\font\ottorm=cmr8\font\ottoi=cmmi8\font\ottosy=cmsy8
\font\ottobf=cmbx8\font\ottott=cmtt8\font\ottosl=cmsl8\font\ottoit=cmti8
\font\sixrm=cmr6\font\sixbf=cmbx6\font\sixi=cmmi6\font\sixsy=cmsy6
\font\fiverm=cmr5\font\fivesy=cmsy5\font\fivei=cmmi5\font\fivebf=cmbx5
\def\ottopunti{\def\rm{\fam0\ottorm} \textfont0=\ottorm
\scriptfont0=\sixrm \scriptscriptfont0=\fiverm \textfont1=\ottoi
\scriptfont1=\sixi \scriptscriptfont1=\fivei \textfont2=\ottosy
\scriptfont2=\sixsy \scriptscriptfont2=\fivesy \textfont3=\tenex
\scriptfont3=\tenex \scriptscriptfont3=\tenex \textfont\itfam=\ottoit
\def\it{\fam\itfam\ottoit} \textfont\slfam=\ottosl
\def\sl{\fam\slfam\ottosl} \textfont\ttfam=\ottott
\def\tt{\fam\ttfam\ottott} \textfont\bffam=\ottobf
\scriptfont\bffam=\sixbf \scriptscriptfont\bffam=\fivebf
\def\bf{\fam\bffam\ottobf} \tt \ttglue=.5em plus.25em minus.15em
\setbox\strutbox=\hbox{\vrule height7pt depth2pt width0pt}
\normalbaselineskip=9pt\let\sc=\sixrm \normalbaselines\rm}
\catcode`@=11
\def\footnote#1{\edef\@sf{\spacefactor\the\spacefactor}#1\@sf
\insert\footins\bgroup\ottopunti\interlinepenalty100\let\par=\endgraf
\leftskip=0pt \rightskip=0pt \splittopskip=10pt plus 1pt minus 1pt
\floatingpenalty=20000
\smallskip\item{#1}\bgroup\strut\aftergroup\@foot\let\next}
\skip\footins=12pt plus 2pt minus 4pt\dimen\footins=30pc\catcode`@=12
\newdimen\xshift \newdimen\xwidth \newdimen\yshift
\def\ins#1#2#3{\vbox to0pt{\kern-#2 \hbox{\kern#1
#3}\vss}\nointerlineskip} \def\eqfig#1#2#3#4#5{ \par\xwidth=#1
\xshift=\hsize \advance\xshift by-\xwidth \divide\xshift by 2
\yshift=#2 \divide\yshift by 2 \line{\hglue\xshift \vbox to #2{\vfil #3
\includegraphics{#4.ps} }\hfill\raise\yshift\hbox{#5}}} \def\8{\write13}

\def\V#1{{\,\underline#1\,}}
\def\T#1{#1\kern-4pt\lower9pt\hbox{$\widetilde{}$}\kern4pt{}}
\let\dpr=\partial \let\io=\infty\let\ig=\int
\def\fra#1#2{{#1\over#2}}\def\media#1{\langle{#1}\rangle}\let\0=\noindent
\def\guida{\leaders\hbox to 1em{\hss.\hss}\hfill}
\def\tende#1{\vtop{\ialign{##\crcr\rightarrowfill\crcr
\noalign{\kern-1pt\nointerlineskip} \hglue3.pt${\scriptstyle
#1}$\hglue3.pt\crcr}}} \def\otto{\
{\kern-1.truept\leftarrow\kern-5.truept\to\kern-1.truept}\ }

\def\pagina{\vfill\eject} \let\ciao=\bye

\def\st{\scriptscriptstyle} \def\*{\vskip0.3truecm}
\def\lis#1{{\overline #1}} \def\eg{\hbox{\it
e.g.\ }}  \def\ie{\hbox{\it i.e.\ }} \def\fiat{{}}  \def\FF{{\cal F}}
 
\def\\{\hfill\break} \def\={{ \; \equiv \; }}
  \fiat \def\kk{{\V k}}    \def\kk{{\V k}}
\def\annota#1{\footnote{${}^#1$}}
\ifnum\aux=1\BOZZA\else\relax\fi
\ifnum\tipoformule=1\let\Eq=\eqno\def\eq{}\let\Eqa=\eqno\def\eqa{}
\def\equ{{}}\fi
 
\def\1{\ifnum\mgnf=0\pagina\else\relax\fi}
\def\W#1{#1_{\kern-3pt\lower7.5pt\hbox{$\widetilde{}$}}\kern2pt\,}

\fiat
\vskip.5cm
\centerline{\titolo Equivalence of dynamical ensembles}
\centerline{\titolo and Navier Stokes equations\annota{*}{\rm{\it
mp$\_$arc@math.utexas.edu}, \#96-???, and
{\it chao-dyn@xyz.lanl.gov},\#960????}}
\vskip1.cm
\centerline{\dodiciit G. Gallavotti}
\centerline{Fisica, Universit\`a
di Roma La Sapienza, P.le Moro 2, 00185, Roma, Italia.}
 
\vskip1cm
 
\0{\it Abstract: a reversible version of the Navier Stokes equation is
studied.  A conjecture emerges stating the equivalence between the
reversible equation and the usual Navier Stokes equation.  The latter
appears as a statement of ensembles equivalence in the limit of
infinite Reynolds number, which plays the role of the thermodynamic
limit.} \*
 
\0{\it Keywords}: {\sl Strange attractors, Chaoticity principle, Fluid
mechanics, Navier-Stokes equation}
 
\vskip1.5truecm\numsec=1\numfor=1
 
In reference [GC1] a scheme for applying the chaoticity principle, also
introduced there, to fluid motions was discussed, and in [G5] some
other simple consequences were derived.
 
We begin by remarking that the usual derivation of the NS equations
assumes on an empirical basis that there is a viscous force between
fluid layers flowing at different speeds, and transforming kinetic
energy into heat: it is desirable that such an assumption be justified
microscopically, [Sp].  On the other hand it is difficult to see how a
microscopically reversible equation of motion, non trivially driven by
a non conservative force, could possibly lead to macroscopically
irreversible equations of the NS type, with energy converted into heat
in a stationary fashion.  Unless a dissipation mechanism is {\it a
priori} assumed at a microscopic level.
 
One could be tempted to say that irreversible macroscopic motion arises
because of a mechanism like the one discussed by Grad and Lanford in
their derivation of the Boltzmann equation.  Note however that,
although the Boltzmann equation is certainly irreversible, it seems
that it cannot be used to derive the NS equations in a situation in
which there are non conservative external forces acting on the system,
at least not without extra assumptions besides the ones one is willing
to assume in deriving the Boltzmann equation itself in a system driven
to equilibrium.\annota{1}{Note also that such a derivation will necessarily
involve proving an existence theorem for the Boltzmann equation of a
confined system for a very long (in fact I believe infinite) time and
this is out of reach of the presently known techniques, [La].}

The physical problem that is stressed here, among many, is that the
work performed by the forces on the system will not be dissipated and
the system {\it will heat up} indefinitely, without reaching a
stationary state, hence without reaching a state described by a
stationary probability distribution for the NS equation.

Starting (to fix the ideas) from the reversible microscopic dynamics,
the kind of "extra assumptions" consist essentially in supposing
suitably large ratios between the microscopic time and length scales
and the macroscopic ones: in order to make effectively infinite the time
scale over which heat is created.  If the scale ratios are very large
and their relative values also suitably adjusted then the motions may
be ``well approximated'' by a dissipative equation and reach a
stationary state (very slowly drifting away from itself).  {\it Strictly
speaking the equations will still be reversible} (involving perhaps
also other fields like the temperature field) and will be governed by
approximately defined transport coefficients equal in the average (over
time and space) to the observed ones.

The estimate of the size of the various scale ratios for this picture
to be of practical interest, and the corresponding evaluation of the
errors, is a difficult problem that, as far as I know, has not only no
clear solution but the methods for its solution do not seem to exist
yet, [Sp].

The above remarks also show that the macroscopic dissipative equations
may perhaps be equivalent to reversible equations: the aim of
this paper is to argue about the possibility of notions
like ``non equilibrium ensembles'' and of interpreting the equivalence
statement as a statement about ensembles equivalence.

For the above reasons it will be useful to imagine an idealized
dissipation mechanism. Here we shall consider a dissipation 
due to the interaction of the system with a "thermostat" which
will be modeled by a force acting on the system: the function of such
forces will be that of {\it imposing} a constant energy
dissipation rate.  The force will be determined by Gauss'
least constraint principle, [LA], [W].\annota{2}{
This type of thermostatting force is essentially the well known {\it 
Nos\'e--Hoover} thermostat and it has been studied in particular in 
the fundamental paper [ECM2] which led to the formulation
of the chaotic hypothesis as a reinterpretation of the earlier
Ruelle's principle, [R2], [G2], [GC2].}
 
It is easy to derive the equations of a fluid in such a situation.
We suppose that: 

\0(1) the fluid is in a periodic box $\O$ with side $L$,\\
\0(2) that it is incompressible with density $\r$,\\
\0(3) that the energy dissipation is:
 
$$\e=\h\fra12\ig\sum_{ij}(\dpr_i u_j+\dpr_j u_i)^2\,d\V x
=\h\ig \V \o^2\,d\V x\Eq(1.1)$$
where $\V\o=\V\dpr\wedge\V u$ is the vorticity field and $\h$ is a
parameter equal to the {\it phenomenological} (\ie experimentally
determined) dynamical viscosity: so that $\n=\h\r^{-1}$ is the 
viscosity.
 
Imposing that the fluid is not viscous but subject to the constraint that
\equ(1.1) is a constant of motion, implies that the Euler equations for
the velocity and pressure fields $\V u,p$ are modified into:
 
$$\dot{\V u}+\W u\cdot\W \dpr \,\V u=-\fra1\r\V\dpr p+\V g+\a\D\V u,
\qquad\V\dpr\cdot\V u=0\Eq(1.2)$$
where $\D$ is the Laplace operator and $\a$ is {\it not} the viscosity
but rather it is the multiplier necessary to impose the constraint that
\equ(1.1) is an integral of motion.  A simple computation shows that:
 
$$\a(\V u)=\fra{\ig\big(\V\dpr\wedge
\V g\cdot\V\o+ \V\o\cdot \,(\W \o\cdot\W \dpr\V
u)\big)\,d\V x}{\ig\V\o^2\,d\V x}\Eq(1.3)$$
This has odd symmetry in $\V u$, so that \equ(1.2) is {\it reversible}:
if $V_t$ is the flow describing the equation solution
(so that $t\to V_t \V u=\V u(t)$ is the solution with initial data $\V
u$) then the transformation $i:\V u\to -\V u$ {\it anticommutes} with the
time evolution $V_t$:
 
$$i\,V_t\,=\,V_{-t}\,i\Eq(1.4)$$
The {\it reversible} equation \equ(1.2), \equ(1.3) will be called the
Gaussian Navier Stokes equation, or GNS.

{\it A similar equation, with a constraint on the energy contained in
each ``momentum shell'' to be constant} was considered in [ZJ], which is
the first paper in which the idea of a reversible Navier Stokes equation
is advanced and studied. The energy content of each ``momentum shell'' was
fixed to be the value predicted by Kolmogorov theory, [LL].

Existence of global solutions to the equations
\equ(1.2),\equ(1.3) is non trivial even if the initial datum $\V u_0$
is $C^\io$.  In view of the conjecture that follows this might be a
reassuring tribute to the principle of conservation of difficulties.
 
In fact let the velocity field $\V u=\sum_{\V k\ne\V0}\V\g_{\V k} e^{i\V
k\cdot\V x}$ be represented in Fourier series with $\V \g_{\V
k}=\lis{\V \g_{-\V k}}$ and $\V k\cdot\V\g_{\V k}=0$ (incompressibility
condition); here $\V k$ has components that are integer multiples of the
"lowest momentum" $k_0=\fra{2\p}L$. Then consider the equation:
 
$$\eqalignno{
&\dot{\V\g}_{\V k}=-\a \V k^2\V \g_{\V k}
-i\sum_{\V k_1+\V k_2=\V k}
(\V\g_{\V k_1}\cdot\V k_2)\, \P_{\V k}\V
\g_{\V k_2}+\V g_{\V k}\cr
&\a=\a_i+\a_e, \qquad
\a_e=\fra{\sum_{\V k\ne\V0}\kk^2\V g_{\V k}\cdot
\lis{\V \g}_{\V k}}{\sum_{\V k}\V k^2|\V \g_{\V k}|^2}
               &\eq(1.5)\cr
&\a_i=\fra{-i\sum_{\V k_1+\V k_2+\V k_3=\V0}
\V k_3^2\,(\V\g_{\V k_1}\cdot\V k_2)\,(\V \g_{\V k_2}\cdot\V \g_{\V
k_3})}{\sum_{\V k}\V k^2|\V \g_{\V k}|^2}\cr}$$
where the $\V k$'s take only the values $0<|\V k|<K$ for some {\it
momentum cut--off} $K>0$ and $\P_{\V k}$ is the orthogonal projection on
the plane orthogonal to $\V k$. This is an equation that defines a
"truncation on the momentum sphere $K$ of the GNS equations".  The
actual GNS equations \equ(1.2), \equ(1.3) have the same form with
$K=+\io$.

For simplicity we suppose that the mode $\V k=\V 0$ is {\it absent}, \ie
$\V\g_{\V0}=\V0$: this can be done if, as we suppose, the external force
$\V g$ does not have a zero mode component (\ie if it has zero average).
 
Calling $V_t^K\V u$ the solution to the equation \equ(1.5) with initial
datum $\V u\in C^\io$, and with components with mode $\V k$ greater than
$K$ set equal to zero, it should follow that $V^K_t\V u\tende{K\to\io}
V_t\V u$ uniformly in $\V x$ and in $t$ for $t$ in any bounded interval.
This is however {\it not} so easy and in fact I do not know whether
\equ(1.5) admits a global solution that can be constructed in this way.
\annota{3}{Examining the theory of Leray, [L], for the NS equations one
realizes that the ``only'' difficulty there was the absence of a uniform
{\it a priori estimate} on the total vorticity $\ig
\V\o^2\,d\V x$: in the equation \equ(1.2),\equ(1.3) the uniform
boundedness of the total vorticity is simply automatic (as it is a
constant of motion). But the theory of Leray also made essential use of
the constancy of the viscosity coefficient, or more precisely of its
positivity and boundedness away from $0$, see [G1]. The latter property
is false for the coefficient $\a$ which, by reversibility,
takes values of either sign (although, see below, we expect that it is
``more often'' positive than negative on "typical" motions).}
 
For the purpose of comparison with the NS equation we shall therefore
use the equations \equ(1.5) and compare them with the corresponding
truncation of the NS equation (with {\it constant} viscosity $\n$)
subject to the non conservative field $\V g$:
 
$$\dot{\V u}_{\V k}=-\n \V k^2\V \g_{\V k} -i\sum_{\kk_1+\V
k_2=\V k} {\W\g}_{\V k_1}\kern-2mm\cdot\W k_2\, \P_{\V k}\V
\g_{\V k_2}+\V g_{\V k}\Eq(1.6)$$
with the same $K$.  We shall take $K$ very large, fixing it once and for
all and dropping it from the places where it would belong as a label (to
simplify the notations). 

In order that the resulting cut--off equations be physically acceptable,
and supposing that $\V g_\kk\ne\V0$ only for $|\kk|\sim k_0$, one shall
have to fix $K$ much larger than the {\it Kolmogorov scale}
$K=(\e_k\n^{-1})^{1/4}$, where $\e_k$ is the average dissipation of the
solutions to \equ(1.6) with $K=+\io$, determined on the basis of
heuristic dimensional considerations by $\e_k\sim \sqrt{|\V g|^3L}$:
see [LL].

We shall also call the truncated equations \equ(1.5),\equ(1.6) 
respectively GNS and NS equations omitting the "truncated": below we 
always refer to the truncated equations, unless otherwise stated.

The solutions of the equations \equ(1.6) with initial datum $\V u$ will
be denoted $V^{ns}_t\V u$.  And we shall admit that there is a unique
stationary distribution $\m_{ns}$ describing the statistics of all
initial data $\V u$ that are randomly chosen with a "Liouville
distribution", \ie with a distribution $\m_0(d\V \g)$ proportional to
the volume measure $\prod_{|\V k|<K} d \V\g_{\V k}$.
 
This means that given ``any observable" $F$ on the phase space $\FF$ (of
the velocity fields with momentum cut--off $K$) it is:
 
$$\lim_{T\to\io}\fra1T\ig_0^T F(V^{ns}_t \V \g) dt=
\ig_\FF F(\V \g') \m_{ns}(d\V \g')\Eq(1.7)$$
for all choices of $\V \g$ except a set of zero Liouville measure.
The distribution $\m_{ns}$ will be called the SRB distribution for the
Eq. \equ(1.6) (see the "zero-th law" in [UF], [GC2]).
 
We make the corresponding assumptions for the solutions of the cut--off
GNS equations \equ(1.5) (with the same cut--off) and the relative flow
$V^{gns}_t$; we call $\m_{gns}$ the associated SRB distribution, [R2], [ER].
 
In particular we shall study the observable $\s(\V\g)$ that we call the
{\it entropy production rate} and that is given by the divergence of the
r.h.s. of Eq. \equ(1.5).

If $D_K$ is the number of modes $\kk$ with $0<|\kk|<K$ then the number
of (independent) components of $\{\V\g_\kk\}$ is $2D_K$ and, see
\equ(1.5), setting $\lis D_K=\sum_{|\V k|<K}2\V k^2$, one finds:

$$\s=2\lis D_K \, (\a_i+\a_e)+\lis\a_i-\lis\a_e=2 \lis
D_K\a+\lis\a_i-\lis\a_e\Eq(1.8)$$
where $\lis \a_i,\lis\a_e$ are suitably defined: \eg
$$\lis\a_e =\fra{\sum_{\V k} \V k^4\lis{\V g}_{\V k}\cdot\V\g_\kk}
{\sum_{\V k}\V k^2
|\V \g_{\kk}|^2}-
2\fra{\big(\sum_{\V k} \V k^2\lis{\V g}_{\V k}\cdot\V\g_\kk\big)
\big(\sum_{\V k} \V k^4\V\g_\kk^2\big)}{\big(\sum_{\V k}\V k^2
|\V \g_{\kk}|^2\big)^2}\Eq(1.9)$$
so that $\s\simeq 2\lis D_K\a$.

We call $\s_{ns}, \e_{ns}$ the time averages of $\s$ and of
$\r\n L^3\sum_{\V k}|\V k|^2|\V \g_{\V k}|^2$ (total vorticity, see
\equ(1.1)) with respect to the distribution $\m_{ns}$. As mentioned 
we expect, on the basis of dimensional analysis, that
$\e_{ns}\sim\e_k=\sqrt{|\V g|^3L}$, see [LL]).  

{\it Correpondingly} we consider the solutions of the GNS equations
with total vorticity $\e_{gns}\=\e=\e_{ns}$ 
and call $\s_{gns}$ the
average of $\s$ with respect to $\m_{gns}$.
 
The $H$--theorem of Ruelle, [R3], tells us that
$\s_{gns}\ge0$, and $\s_{gns}>0$ if the distribution
$\m_{gns}$ is not proportional to the volume measure. Here we shall
suppose that this is the case if $\V g\ne\V0$: at least if the force
field is such that it generates a chaotic motion (\ie a motion with at
least one positive Lyapunov exponent) for all but a set of zero volume
initial data. In this situation:
\*
\0{\sl Conjecture ("equivalence of dynamical ensembles"):} {\it Given
$\e\=\e_{gns}$ suppose that the parameter $\n$ in the NS equation
is adjusted so that $\e_{ns}=\e_{gns}$ then the
averages $\s_{gns}, \s_{ns}$ of $\s$ with respect to $\m_{gns}$ and
with respect to $\m_{ns}$ and, in fact, the averages of all
(reasonable) observables coincide, up to corrections $O(R^{-1})$,
$R=\e^{1/3}L^{4/3}\n^{-1}$.} \*
 
We recognize in this conjecture a statement very analogous to the
familiar statements on the equivalence of thermodynamic ensembles, with
the thermodynamic limit replaced by the limit $R\to\io$ of infinite
Reynolds number. The paper [ZJ] contains various statements that can be
seen as steps towards the formalization given above of the general
conjecture.
 
It can be weakened very much for the purposes of possible applications:
we state it in the above very strong form to stress an extreme
thought. Applications will be envisaged elswhere.

On heuristic grounds, it would be justified if one did accept that the
entropy creation rate reaches its average on a time scale that is fast
compared to the hydrodynamical scales.  The coefficient $\a\simeq(2\lis
D_K)^{-1}\s$, see \equ(1.8), would be confused with its average
$\media{\a}_{gns}$ and {\it identified with the viscosity} constant
$\n$.

In this way the GNS and the NS equations would be equivalently good: both
being the macroscopic trace of two {\it equivalent microscopic
dissipation mechanisms}: one explicitly specified by the gaussian
constraint of constant dissipation and the other {\it unspecified but
phenomenologically modeled} by a constant viscosity.

The interest of the conjecture is that it allows us to deduce properties 
of the "usual" NS equation from properties of the GNS equation: the 
latter, being reversible, can be studied via the {\it chaotic 
hypothesis} of [GC1], [GC2] which leads to non trivial predictions like the 
fluctuation theorem of [GC1].

\*

\0{\it Acknowledgements:} I am grateful to 
F.  Bonetto, G. Gentile, C.  Liverani, V. Mastropietro for many
discussions and suggestions.  This work is part of the research program
of the European Network on: "Stability and Universality in Classical
Mechanics", \# ERBCHRXCT940460, and it has been partially supported also
by CNR-GNFM and Rutgers University.

\0{\bf References.}
\*
\item{[ECM2]} Evans, D.J.,Cohen, E.G.D., Morriss, G.P.: {\it Probability
of second law violations in shearing steady flows}, Physical Review
Letters, {\bf 71}, 2401--2404, 1993.

\item{[ER]} Eckmann, J.P., Ruelle, D.: {\it Ergodic theory of strange
attractors}, Reviews of Modern Physics, {\bf 57}, 617--656, 1985.
 
\item{[G1]} Gallavotti, G.: {\it Some rigorous results about 3D Navier 
Stokes}, Les Houches 1992 NATO-ASI meeting on ``Turbulence in spatially
extended systems'', ed. by R. Benzi, C. Basdevant, S. Ciliberto;
p. 45--81; Nova Science Publishers, New York, 1993.

\item{[G2]} Gallavotti, G.: {\it Ergodicity, ensembles,
irreversibility in Boltzmann and beyond}, Journal of Statistical
Physics, {\bf 78}, 1571--1589, 1995. See also {\it Topics in chaotic
dynamics}, Lectures at the Granada school, ed. Garrido--Marro, Lecture
Notes in Physics, {\bf 448}, 1995. And, for mathematical details, {\it
Reversible Anosov diffeomorphisms and large deviations.},
Ma\-the\-ma\-ti\-cal Physics Electronic Journal, {\bf 1}, (1), 1995,
(http:// www. ma. utexas. edu/ MPEJ/ mpej.htlm).
 
\item{[G3]} Gallavotti, G.:{\it Chaotic hypothesis: Onsager reciprocity
and fluctuation--dis\-si\-pa\-tion theorem}, in {\it
mp$\_$arc@math.utexas.edu}, \#95-288, 1995. In print in Journal of
Statistical Physics, 1996.
 
\item{[G4]} Gallavotti, G.: {\it Chaotic principle: some applications
to developed turbulence}, Archived in {\it mp$\_$arc@math.utexas.edu},
\#95-232, and {\it chao-dyn@ xyz.lanl.gov}, \#9505013.
 
\item{[G5]} Gallavotti, G.: {\it Extension of Onsager's reciprocity to 
large fields and the chao\-tic hypothesis}, http:// chimera. roma1. infn. it,
1996, or mp$\_$ arc 96-??; or chao-dyn 9603003.

\item{[GC1]} Gallavotti, G., Cohen, E.G.D.: {\it Dynamical ensembles in
nonequilibrium statistical mechanics}, Physical Review Letters,
{\bf74}, 2694--2697, 1995.
 
\item{[GC2]} Gallavotti, G., Cohen, E.G.D.: {\it Dynamical ensembles in
stationary states}, in print in Journal of Statistical Physics, 1995.
 
\item{[L] } Leray, J.: {\it Sur le mouvement d' un liquide
visqueux remplissant l' espace}, Acta Mathematica, 63, 193- 248, 1934.
 
\item{[La]} Lanford, O.: {\it Time evlution of large classical
systems}, in "Dynamical systems, theory and applications", p.  1--111,
ed.  J.  Moser, Lecture Notes in Physics, {\bf 38}, 1974, Springer
Verlag.

\item{[LA]} Levi-Civita, T., Amaldi, U.: {\it Lezioni di Meccanica
Razionale}, Zanichelli, Bo\-lo\-gna, 1927 (re\-prin\-ted 1974), volumes
$I,II_1,II_2$.
 
\item{[LL]} Landau, L., Lifshitz, V.: {\it M\'ecanique des fluides}, MIR, 
Moscow, 1966.
 
\item{[R2]} Ruelle, D.: {\it Chaotic motions and strange attractors},
Lezioni Lincee, notes by S.  Isola, Accademia Nazionale dei Lincei,
Cambridge University Press, 1989; see also: Ruelle, D.: {\it Measures
describing a turbulent flow}, Annals of the New York Academy of
Sciences, {\bf 357}, 1--9, 1980.
 
\item{[R3]} Ruelle, D.: {\it Positivity of entropy production in non
equilibrium statistical mechanics}, IHES preprint, 1995.

\item{[Sp]} Spohn, H.: {\it Large scale dynamics of interacting
particles}, Springer--Verlag, 1991.
 
\item{[UF]} Uhlenbeck, G.E., Ford, G.W.: {\it Lectures in Statistical
Mechanics}, American Mathematical society, Providence, R.I.,
pp. 5,16,30, 1963.
 
\item{[W]} Wittaker, E.T.: {\it A treatise on the analytical dynamics 
of particles \& rigid bodies}, Cambridge U. Press, 1989 (reprinted).

\item{[ZJ]} Zhen-Su She, Jackson, E.: {\it Constrained Euler system for 
Navier--Stokes turbulence}, Physical Review Letters, {\bf 70},
1255--1258, 1993.
\*
\0{\it Internet access:
\*
All the Author's quoted preprints can be found and freely downloaded
(latest postscript version including corrections of misprints and
errors) at:

\centerline{\tt http://chimera.roma1.infn.it}

\0in the Mathematical Physics Preprints page.}

\ciao